\documentclass[%
 reprint,
 amsmath,amssymb,
 aps,
]{revtex4-2}

\usepackage{graphicx}
\usepackage{dcolumn}
\usepackage{bm}

\usepackage{soul}
\usepackage{cancel}
\usepackage{ulem}
\usepackage[dvipsnames]{xcolor}
\begin{document}

\preprint{APS/123-QED}

\title{Kerr-Nonlinearity Assisted Exceptional Point Degeneracy in a Detuned $PT$-Symmetric System}

\author{Shahab Ramezanpour}
\email{shahab.ramezanpour@utoronto.ca}
\affiliation{The Edward S. Rogers Department of Electrical and Computer Engineering, University of Toronto}

\date{\today}

\begin{abstract}
Systems operating at exceptional points (EPs) are highly responsive to small perturbations, making them suitable for sensing applications. Although this feature impedes the system working exactly at an EP due to imperfections arising during the fabrication process. We propose a fast self-tuning scheme based on Kerr nonlinearity in a coupled dielectric resonator excited through a waveguide placed in the near-field of the resonators. We show that in a coupled resonator with unequal Kerr-coefficients, initial distortion from EP regime can be completely compensated. It provides an opportunity to reach very close to the EP in a coupled resonator with detuned resonant frequencies via tuning the intensity of the incident wave. Using time-modulation of the incident wave in nonlinear systems to control both the gain or loss, and resonant frequencies can be a possible approach to fully control the parameters close to an EP.
\end{abstract}

\maketitle

\section{Introduction}
\textit{Exceptional Points} (EPs) are degeneracy in non-Hermitian systems that have attracted many interests in recent years due to their broad applications in photonics ~\cite{EP_opt_pho,PT_WGM,EP_enhance_sen,sen_high_EP}. In practice, achieving the EP-regime can be challenging, since EP-based systems are highly responsive to imperfections arising in the fabrication process. Though in many applications, it is not required to  work exactly at an EP, it is desirable to have a way to control the system's status. Therefore, a subsidiary mechanism is here investigated to tune these systems close to an EP. Secondary mechanisms can be, such as a heating scheme to tune the resonant frequencies of coupled resonators ~\cite{PT_WGM,sen_high_EP} or a nanoparticle as a perturbation to tune the (radiation) loss and resonant frequency of resonators ~\cite{loss_lasing,phonon_lasing}. For example, ~\cite{GEP} derives general conditions of EP in a perturbed coupled resonator which contains one or two nanoparticles, and a highly chiral EP is achieved in a coupled resonator perturbed by a nanoscatterer ~\cite{CEP}. Another interesting scheme is using nonlinearity in the coupled resonators. The eigenvalue analysis of such  a system is performed in ~\cite{tuning_EP}, which shows that using a coupled resonator with unequal Kerr-nonlinearities can exhibit an EP for the resonators with unequal resonant frequencies.  

To have an EP in coupled resonators, specific relations between the system parameters, including gain or loss, coupling rate, etc. are required. However, instead of using explicit gain or loss, in ~\cite{PT_time_mod} time-modulation is applied as an equivalent mean to produce gain or loss. Virtual PT-symmetric systems ~\cite{virtual_PT} can have applications in critical coupling in high-Q resonators  ~\cite{virtual_coupling} and pulling force for a passive resonant object of any shape and composition ~\cite{virtual_force}. Furthermore, instead of using gain or loss in a two-resonator system, an EP degeneracy has been realized in a single resonator with a time-periodic component ~\cite{linear_time_mod,linear_time_mod2}, time modulation of a single resonator ~\cite{sen_time_mod} or a distributed parameter in a transmission line ~\cite{time_mod_TL}. Exceptional degeneracies are also observed in  waveguides loaded with discrete gain and loss elements ~\cite{w_g_l}, and in lossless/gainless coupled-resonators optical waveguides ~\cite{CROW}.  
Another way to get EPs is based on two resonators coupled by a gyrator ~\cite{gyrator}. 

Instead of material properties, shape and dimension of resonators can be manipulated to achieve desired optical responses, tune their energy spectra, and create specific resonances related to such as EP and Bound State in Continuum (BIC). By engineering the height and radius of a cylinder with a high refractive index, high-Q supercavity mode can be achieved by realizing the regime of BIC ~\cite{supercavity, supercavity_obs}. This structure has shown a considerable enhancement of second harmonic generation ~\cite{res_NL}. The supercavity mode as well as EP can also be realized in a deformed shape of a cylindrical resonator ~\cite{EP_deformed_cylinder}. By manipulating the shape of a semiconductor particle, near unity absorption can be achieved in the broadband optical spectra ~\cite{spie}. Meanwhile, energy spectra as well as whispering gallery modes of semiconductor quantum dot can be controlled in a lateral and corrugated shape ~\cite{MRX}. A deformed disc enables the light to travel between different angular momenta, which can effectively enhance third-harmonic generation ~\cite{chaos_momentum}.EP can be achieved in a deformed disc resonator through the coupling of modes with different angular momenta. ~\cite{EP_deformed,EP_deformed_higher}.

Among of several tunable mechanisms such as material properties and shape of resonators, utilizing Kerr-nonlinearity in resonators can be privileged considering its efficient and fast tuning properties by input power. Incorporating both nonlinearity and non-Hermiticity is intriguing and new functionalities can be emerged such as tuning topological insulators ~\cite{tuning_topology,tuning_topology2,tuning_topology3} and transition from PT-symmetric to broken PT-symmetric ~\cite{transition_NL,bistable_NL} phase. Besides, nonlinearity may provide a suitable platform to improve the sensitivity of EP-based systems in the presence of Quantum noise ~\cite{EP_robust}. 

The eigenvalue of a nonlinear non-Hermitian system shows that the EP can be tuned with respect to a detuning parameter related to resonant frequencies ~\cite{tuning_EP}. 

In contrast to ~\cite{tuning_EP}, which is based on eigenvalue analysis and time-independent coupled mode theory, here, a full-wave numerical simulation in Comsol Multiphysics ~\cite{comsol} is examined within a realistic system. Besides, time-dependent coupled-mode theory is used to study the behavior of the system around EP in the temporal domain. To apply a detuning, a coupled ring resonator is taken into consideration, with the assumption that their radii differ slightly. It is assumed that one of the resonators has a Kerr-type nonlinearity. Instead of changing the nonlinear coefficients (as in ~\cite{tuning_EP}), we change the power of the incident wave to tune the nonlinear strength. Interestingly, for a given incident power, the transmission is maximum (associated with EP degeneracy); below and beyond this power, the transmission coefficient significantly drops, in accordance with our theoretical prediction.

The time domain behavior of the system around EP, which involves detuning and nonlinearity, is rarely studied. Here, the time evolution of the aforementioned structure is studied. The interplay between nonlinearity and detuning shows the shifting of the system to PT-Symmetric and Broken PT-Symmetric regime, as well as the compensation of detuning with nonlinearity in time domain. Within this framework, we could investigate the stability analysis of the system incorporating detuning and nonlinearity, as well as its intriguing physics at the boundary between stable and unstable regimes. This work not only establishes an experimental framework for studying EP in nonlinear detuned coupled resonators, but it also paves the way for basic research on the ways in which nonlinearity and detuning interact to change the system's state while encircling EP.

\section{Theoretical Framework of two coupled Kerr resonators}
Due to imperfections, which may arise in the fabrication process, achieving an EP regime at which both real and imaginary parts of eigenvalues degenerate is challenging. Although, there are some methods to alleviate these imperfections, such as using a heating scheme to tune the resonant frequencies of the resonators and employing a scatterer in the vicinity of a resonator to tune both its loss and resonant frequency, they either can be a slow process, in the former case ~\cite{PT_WGM}, or need a fine positioning and sizing of the scatterer in the latter case ~\cite{loss_lasing}. Here, we propose an efficient fast-process self-tuning scheme using resonators containing different Kerr materials, in which imperfection in the resonant frequencies can be compensated through tuning the intensity of the incident wave. This process is shown schematically in Fig. 1, in which the difference of the resonant frequencies, which appears in eigenmodes $a_-$ and $a_+$, is compensated by increasing the power of the incident wave through the tapered waveguide coupled to the resonators. 
\begin{figure}
\centering
{\includegraphics[width=.45\textwidth]{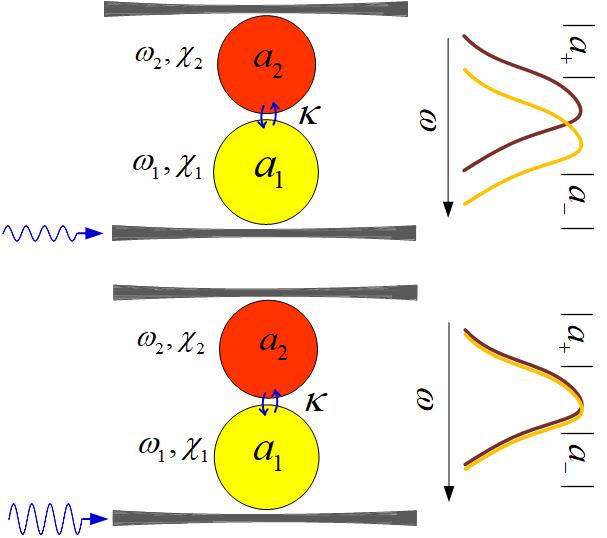}}
\caption{Tuning a coupled resonator with different resonant frequencies to the EP via intensity of the incident wave}
\end{figure}

The equation of motion for mode amplitudes of a dimer made of two resonators with different Kerr nonlinearities, where one of them is coherently driven, can be described by a generalized Gross-Pitaevskii (GP) equation ~\cite{asymmetric_NL,tuning_EP} as
\begin{equation}
i\frac{d}{dt}\left( \begin{matrix}
   {{\psi}_{1}}  \\
   {{\psi}_{2}}  \\
\end{matrix} \right)=H\left( \begin{matrix}
   {{\psi}_{1}}  \\
   {{\psi}_{2}}  \\
\end{matrix} \right)
-\left( \begin{matrix}
   {{S}_{1}}  \\
   0  \\
\end{matrix} \right)
\end{equation}
with
\begin{equation}
H=\left( \begin{matrix}
   \omega_0-i\delta_1+\chi_1|\psi_1|^2  & \kappa_{12}   \\
   \kappa_{21}  & \omega_0+i\delta_2-\epsilon+\chi_2|\psi_2|^2  \\
\end{matrix} \right)\nonumber
\end{equation}
where $\omega_0$ is resonant frequency of the uncoupled resonators, $\delta_1$ is dissipation on the first resonator, $\delta_2$ is gain on the second resonator, $\epsilon$ is detuning of the resonant frequencies, $S_1$ is excitation and $\chi_1$ and $\chi_2$ are the nonlinear Kerr coefficients of the first and second resonators, respectively. According to the energy conservation in the system, the coupling strengths should satisfy the relation $\kappa_{21}=\kappa^*_{12}$. Therefore, we write $\kappa_{12}=\kappa_{21}=\kappa$ for a real value of the coupling strengths.

Under a continuous monochromatic excitation $S_1=s_1e^{-i\omega t}$, by using the ansatz $\psi_j=a_je^{-i\omega t}$, ($j=1,2$), and $\frac{d\psi_j}{dt}=-i\omega a_j e^{-i\omega t}+\frac{da_j}{dt}e^{-i\omega t}$, time-independent GP 
\begin{subequations}
\begin{align}
 & \omega {{a}_{1}}=(\omega_0-i\delta +{{\chi}_{1}}|{{a}_{1}}{{|}^{2}}){{a}_{1}}+\kappa {{a}_{2}}-{{s}_{1}} \\ 
 &\omega {{a}_{2}}=\kappa {{a}_{1}}+(\omega_0+i\delta-\epsilon +{{\chi}_{2}}|{{a}_{2}}{{|}^{2}}){{a}_{2}} 
\end{align}
\end{subequations}
can be obtained by considering slowly varying wave approximation ~\cite{asymmetric_NL,stability,PT_WGM}, where the coupled resonators with balanced gain and loss ($\delta_1=\delta_2=\delta$) is assumed.

The mode amplitudes are calculated from eq. (2) as
\begin{subequations}
\begin{align}
  & {{a}_{1}}=\frac{(-\Delta +i\delta-\epsilon +{{\chi}_{2}}|{{a}_{2}}{{|}^{2}}){{s}_{1}}}{(-\Delta-i\delta +{{\chi}_{1}}|{{a}_{1}}{{|}^{2}})(-\Delta+i\delta-\epsilon +{{\chi}_{2}}|{{a}_{2}}{{|}^{2}})-{{\kappa}^{2}}} \\ 
 & {{a}_{2}}=\frac{-\kappa {{s}_{1}}}{(-\Delta-i\delta +{{\chi}_{1}}|{{a}_{1}}{{|}^{2}})(-\Delta+i\delta-\epsilon +{{\chi}_{2}}|{{a}_{2}}{{|}^{2}})-{{\kappa }^{2}}}  
\end{align}
\end{subequations}
where $\Delta=\omega-\omega_0$. From Eq. 3, we deduce that for the linear case, $\chi_1=\chi_2$=0, and at the EP condition associated to PT symmetry, $\epsilon=0$, $\delta=\kappa$, the field amplitudes tend to infinity when $\Delta \to 0$.  Therefore, small perturbation can significantly affect the mode amplitudes, and we can observe the effect of detuning in the field modes. 

To avoid divergence of $a_1$ and $a_2$ in the linear case, and better convergence in the nonlinear one, ($\delta$,$\kappa$) are considered to be equal to ($1,1-10^{-3}$), i.e., the system is very close, but not exactly at, the EP. After numerically solving eq. (2) ~\cite{tuning_EP}, Fig. 2(a) shows the field amplitudes for the linear case and without detuning parameter (i.e., $\epsilon=0$), whereas Fig. 2(b) shows the modes for the linear case with a small detuning of $\epsilon=0.004$. The  applied source, $s_1$ is equal to 0.01. 
Comparing Figure 2(b) (perturbed) to Fig. 2(a) (unperturbed), we can observe that  amplitude (linewidth) of the field modes decrease (increase) considerably.
\begin{figure}
\centering
{\includegraphics[width=.49\textwidth]{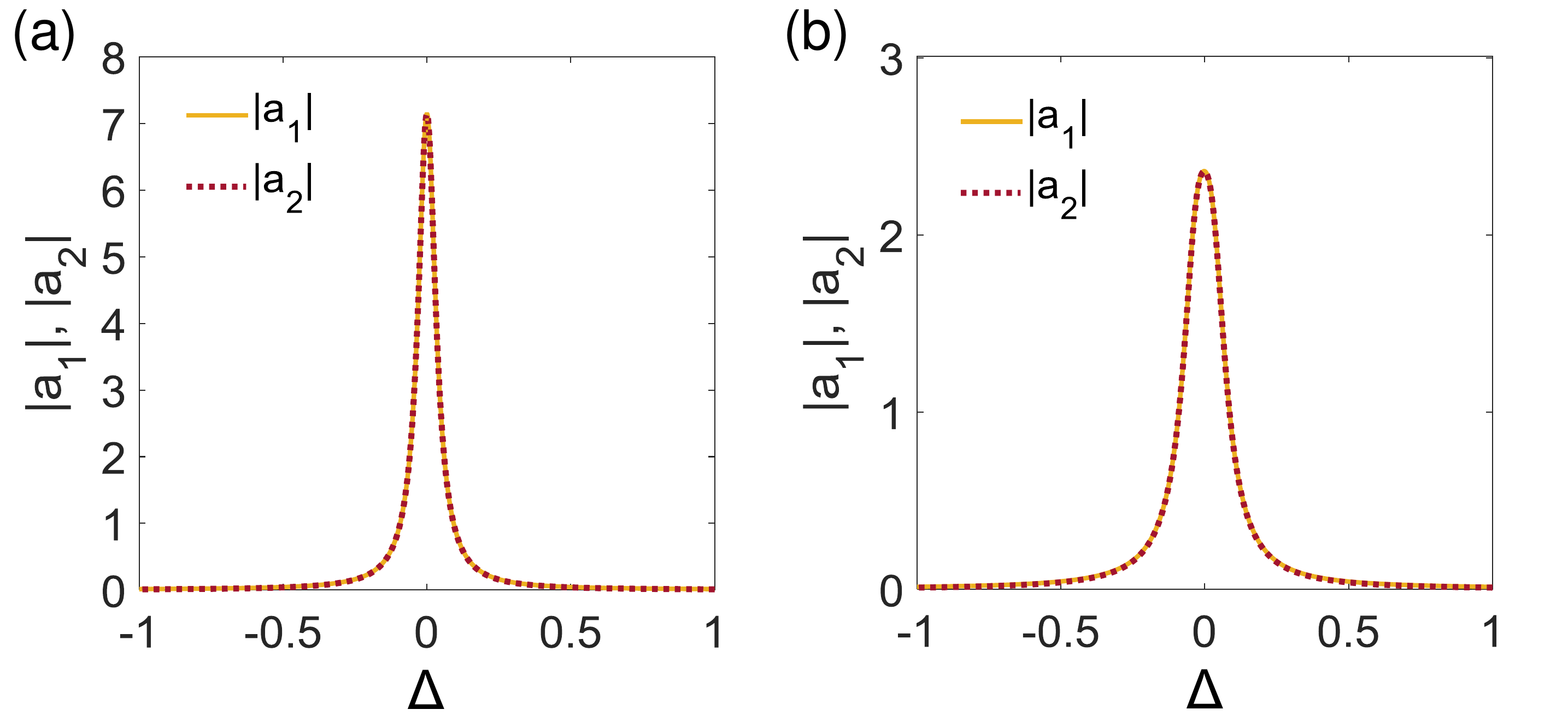}}
\caption{Field amplitudes  for $(\delta,\kappa)$=($1,1-10^{-3}$), $\chi_2=0$ and (a) $\epsilon=0$ (b) $\epsilon=0.004$.}
\end{figure}

In order to compensate for the detuning with a nonlinearity, we consider a specific example with detuning $\epsilon=0.002$, and calculate the field modes versus nonlinearity $\chi_2$, for $\Delta=0$ (Fig. 3(a)). Figure 3(a) shows that the maximum of the modes occurs at  $\chi_2=4\times10^{-5}$, and Fig. 3(b) shows the field modes versus frequency for $\epsilon=0.002$ and $\chi_2=4\times10^{-5}$. The detuned system compensated with a nonlinearity is represented by Fig. 3(b), where the field amplitudes are almost identical to those in Fig. 2(a) (linear system without detuning parameter), but with slightly reduced linewidth. Figure 3(c) shows a 3D plot of the $a_1$ field  amplitude versus $\chi_2$ and $\Delta$ for $\epsilon=0.002$, which shows that the maximum occurs at $\Delta=0$ and $\chi_2=4\times10^{-5}$. The decaying above of the $\chi_2=4\times10^{-5}$ is much faster than below of $\chi_2=4\times10^{-5}$.
\begin{figure}
\centering
{\includegraphics[width=.49\textwidth]{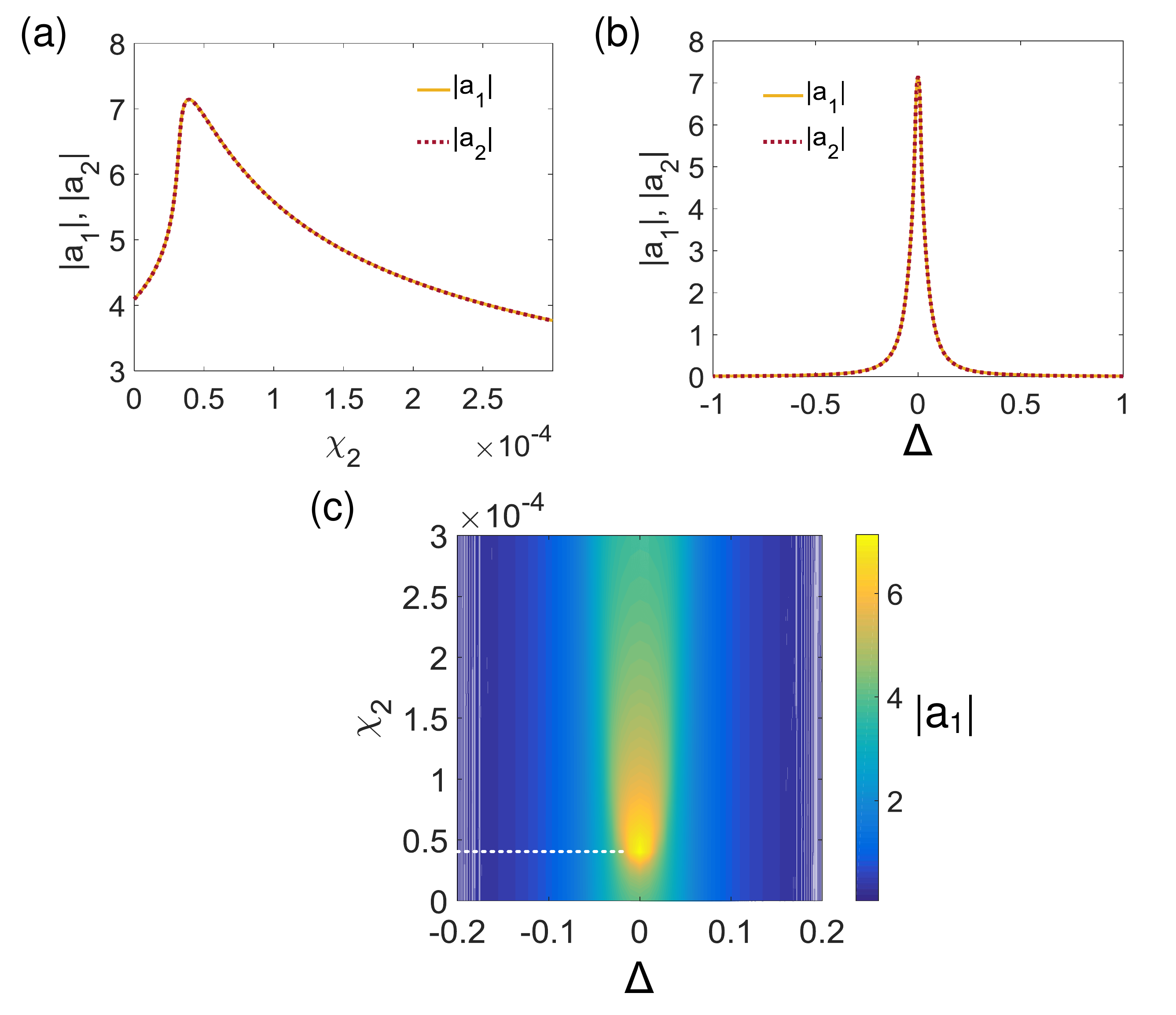}}
\caption{Mode amplitudes (a) versus $\chi_2$ for $\epsilon=0.002$ and $\Delta=0$, (b) versus $\Delta$ for $\chi_2=4\times10^{-5}$ and $\epsilon=0.002$, (c) versus $\chi_2$ and $\Delta$ for $\epsilon=0.002$.}
\end{figure}

\section{Numerical Analysis}
The Kerr nonlinearity is an efficient mechanism to tune the resonant frequency of a resonator by setting the power of the incident wave. This tuning is crucial in high-sensitive systems based on EP since small perturbations can considerably affect its performance. We use the finite element method  implemented in Comsol Multiphysics ~\cite{comsol} to analyze a single ring resonator as well as a coupled ring resonator containing a Kerr nonlinearity, in a 2D, z-invariant model. Figure 4(a) shows the $z$-component of electric field, $|E_z|$ of a single resonator coupled to a waveguide, at its resonant wavelength $1.565\,\mu$\text{m}. The radius of the ring is $r_0/\lambda_0=4$ with $\lambda_0=1.55\,\mu$\text{m}, the refractive indices of core and cladding are 2.5 and 1.5, respectively, and the width of the core (including waveguide) is $0.2\,\mu$\text{m}.  Figure 4(b) shows the transmittance $T=|S_{21}|^2$, where $S_{21}$ is the S-parameter relating port 1 to port 2, for two radii: $r_0/\lambda_0=4$ (solid line), $r_0/\lambda_0=4-10^{-3}$ (dashed-dotted line). The case with $r_0/\lambda_0=4-10^{-3}$ contains the Kerr nonlinearity (dotted line), where the nonlinearity is tuned to achieve the resonant frequency equal to the resonator with radius $r_0/\lambda_0=4$. The incident power in the simulation was $P_{in}=1$ W, which lead to a tuning nonlinearity of $\chi=2.6\times10^{-7}\,\text{cm}^2/$\text{W}, where $\chi$ is considered in the refractive index of the nonlinear resonator as
\begin{equation}
    n=n_{\text{core}}+\frac{c\epsilon_0n_{\text{core}}\chi}{2}|E|^2
\end{equation}
with $n_{\text{core}}=2.5$ and $c=3\times10^{8}\,$m/s.
\begin{figure}
\centering
{\includegraphics[width=.49\textwidth]{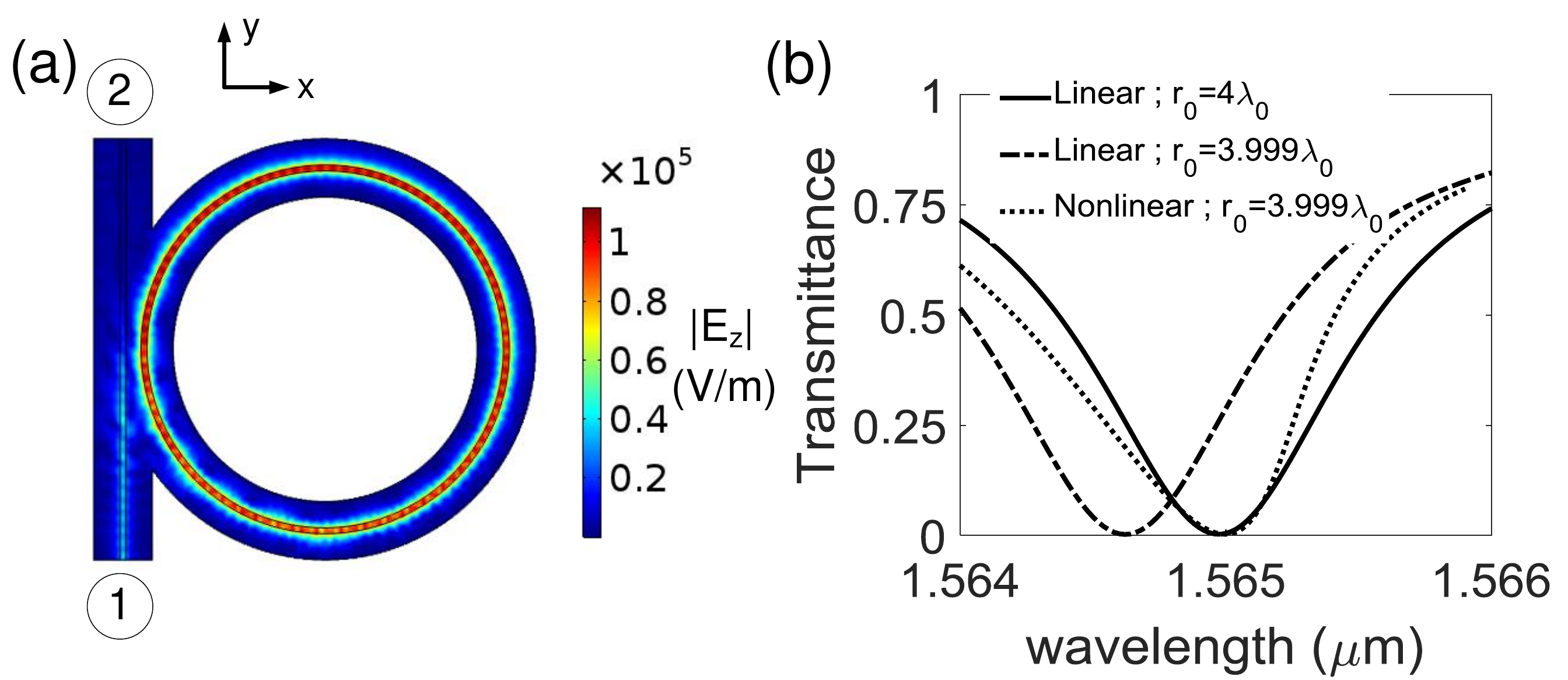}}
\caption{A single resonator without and with Kerr-nonlinearity. (a) Field distribution. (b) Transmittance.}
\end{figure}

In the previous section, we have shown that a detuning in the degenerate resonant frequency of a coupled resonator can be compensated by utilizing Kerr nonlinearity. In Fig. 5, we first tune a coupled resonator with balanced gain and loss to the EP, then we apply a detuning in the radius of one resonator, followed by introducing the nonlinearity to compensate for the detuning.
\begin{figure}[t]
\centering
{\includegraphics[width=.49\textwidth]{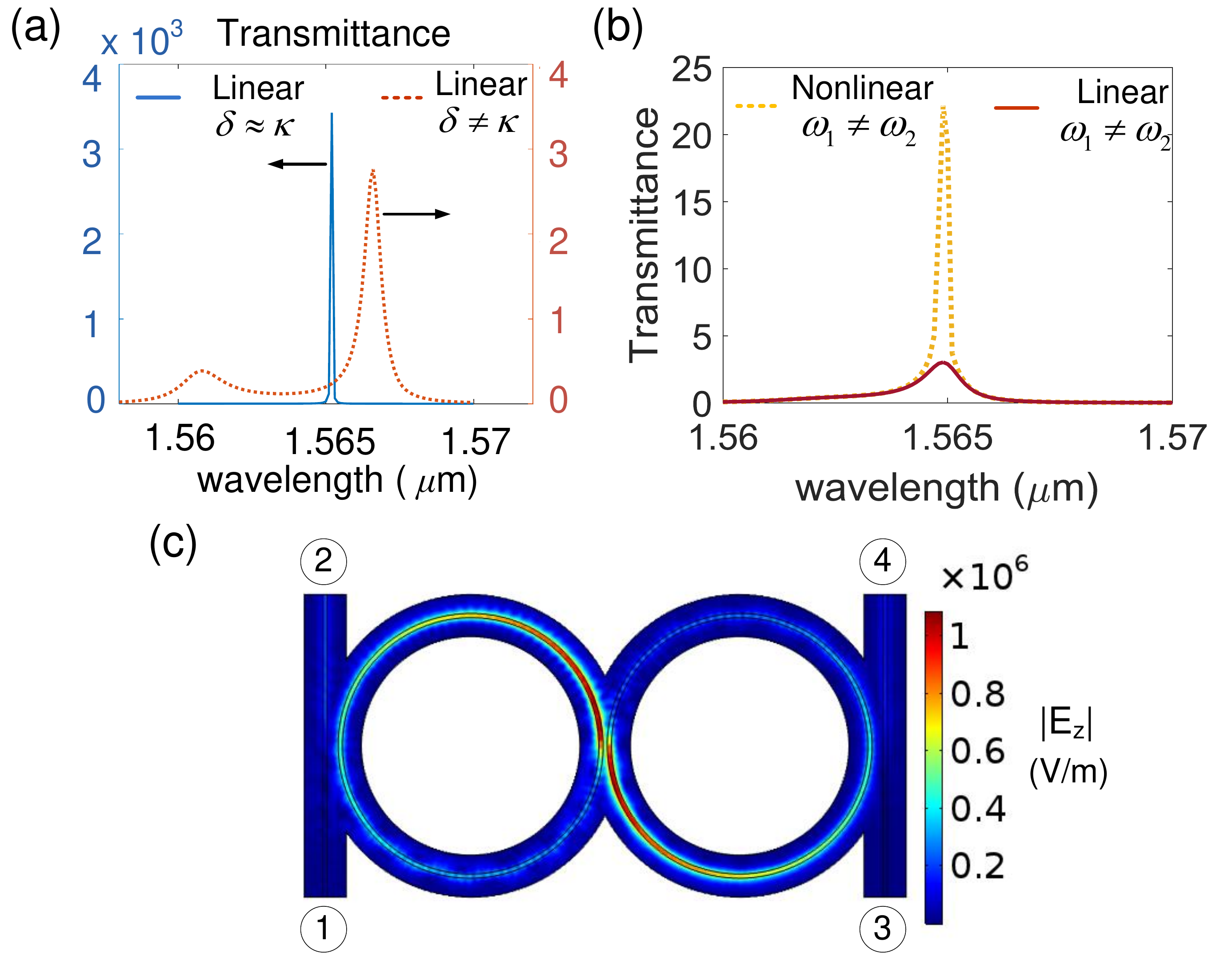}}
\caption{Coupled resonator: (a) Transmittance far from EP and at EP. (b) Transmittance of detuned and nonlinearity assisted coupled resonators. (c) Field distribution at EP.}
\end{figure}

Figure 5(a) shows the transmittance $|S_{41}|^2$ of the coupled resonators with radii $r_{1,2}/\lambda_0=4$, and refractive indices $n_1=2.5-i0.016$, $n_2=2.5+i0.016$ (dotted line) and $n_1=2.5-i0.0174$, $n_2=2.5+i0.0174$ (solid line). The solid line has the characteristic of an EP, with one resonance close to the resonance of a single resonator. Figure 5(b) (solid line) shows the transmittance of the same structure with radii $r_1/\lambda_0=4$, $r_2/\lambda_0=4-10^{-3}$, and the same refractive indices as in Fig. 5(a), in which the magnitude of the transmittance at EP is reduced drastically. To compensate for this detuning, Kerr nonlinearity is considered in the second resonator with 
$\chi=8.07\times10^{-9}\,\text{cm}^2/$\text{W} (dotted line). This value of $\chi$ is found by analyzing the system with different values of $\chi$, and incident power $P_{in}=1$ W. We note that we can fix the value of nonlinearity and find the suitable amount of incident power in the same manner. For large values of $\chi$, the convergence of the numerical simulation could be hardly achieved at an EP, and the number of iterations (a setting in Comsol) is increased to solve this problem. The $|E_z|$ distribution of the coupled resonator containing nonlinearity, at the EP, is shown in Fig. 5(c).

In a realistic configuration, we tune the system to an EP by adjusting the input power. Figure 6(a) and 6(b) shows transmittance $|S_{41}|^2$ and $|S_{21}|^2$, respectively, in which $\chi=8.07\times10^{-9}\,\text{cm}^2/$W  is considered. This amount of Kerr nonlinearity can be obtained in, for instance, an oil-filled cavity ~\cite{kerr}. Analogous to Fig. 3(c), at $P_{in}=1$ W (a default value in Figs. 4 and 5), the transmittance is maximum, while below (above) of this value, transmittance reduces drastically (smoothly).
\begin{figure}
\centering
{\includegraphics[width=.49\textwidth]{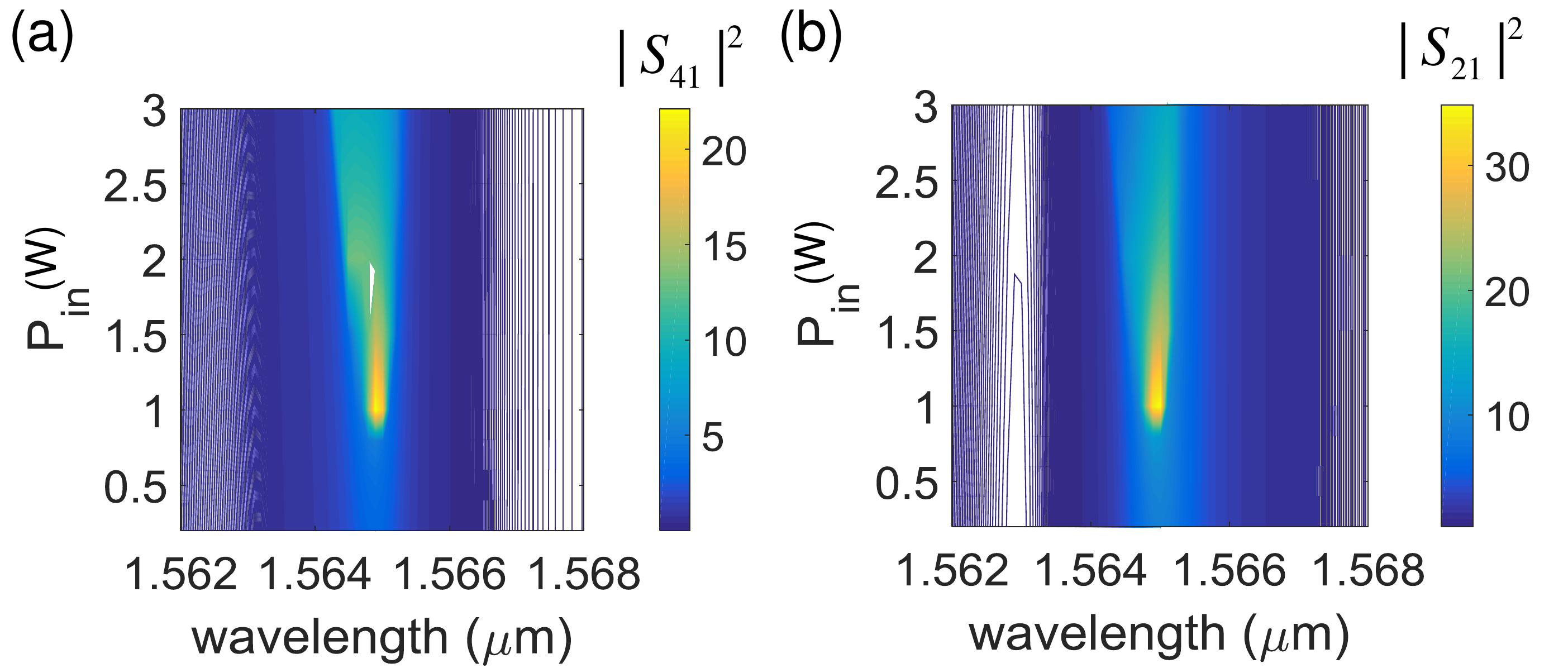}}
\caption{Transmittance calculated (a) from port 4 and (b) port 2 versus input power}
\end{figure}

\section{Dynamics}
In this section, we analyze the dynamic of the system around EP, and solve the time dependent coupled mode equation, eq. (1), numerically. First, a linear system ($\chi_1=\chi_2=0$) with a sinusoidal excitation $S_1=A\text{sin}(\omega t)$, and dimensionless parameters $A=0.01$, $\omega=\omega_0=3$, $\delta_1=-\delta_2=1$, and $\epsilon=0$ is studied. Figure 7, shows time evolution of the field amplitudes $|\psi_j|$ for different value of the coupling around the value $\kappa=1$. We note that for $\kappa>1$, the eigenvalues of the system are entirely real (PT-symmetric region), while for $\kappa<1$ the eigenvalues become complex (broken PT-symmetric region). The boundary of these two regions, $\kappa=1$ has characteristic of EP.  For the coupling $\kappa=1.1$, Fig. 7(a) (left) depicts field amplitudes oscillating by time, and Fig. 7(a) shows the field amplitudes spanning an elliptical route (right). This numerical result is pretty close to analytical one in the supplementary information.

With decreasing the coupling value, $\kappa=1+10^{-3}$, and approaching to the EP, the rate of oscillation would be decreased (Fig. 7(b) (left)), and the span area of $|\hat{\psi}_2|$ vs $|\hat{\psi}_1|$ shrinks. At EP, $\kappa=1$, Fig. 7(c) (left) shows growing of field amplitudes by time, while $|\psi_2|$ vs $|\psi_1|$ shows a linear dependence, $|\psi_1|=|\psi_2|$ (Fig. 7(c) (right)). For $\kappa=1-10^{-4}$, field amplitudes grow by time (Fig. 7(d) (left)), and $|\psi_2|$ vs $|\psi_1|$ shows a linear dependence (Fig. 7(d) (right)).

\begin{figure}
\centering
{\includegraphics[width=.485\textwidth]{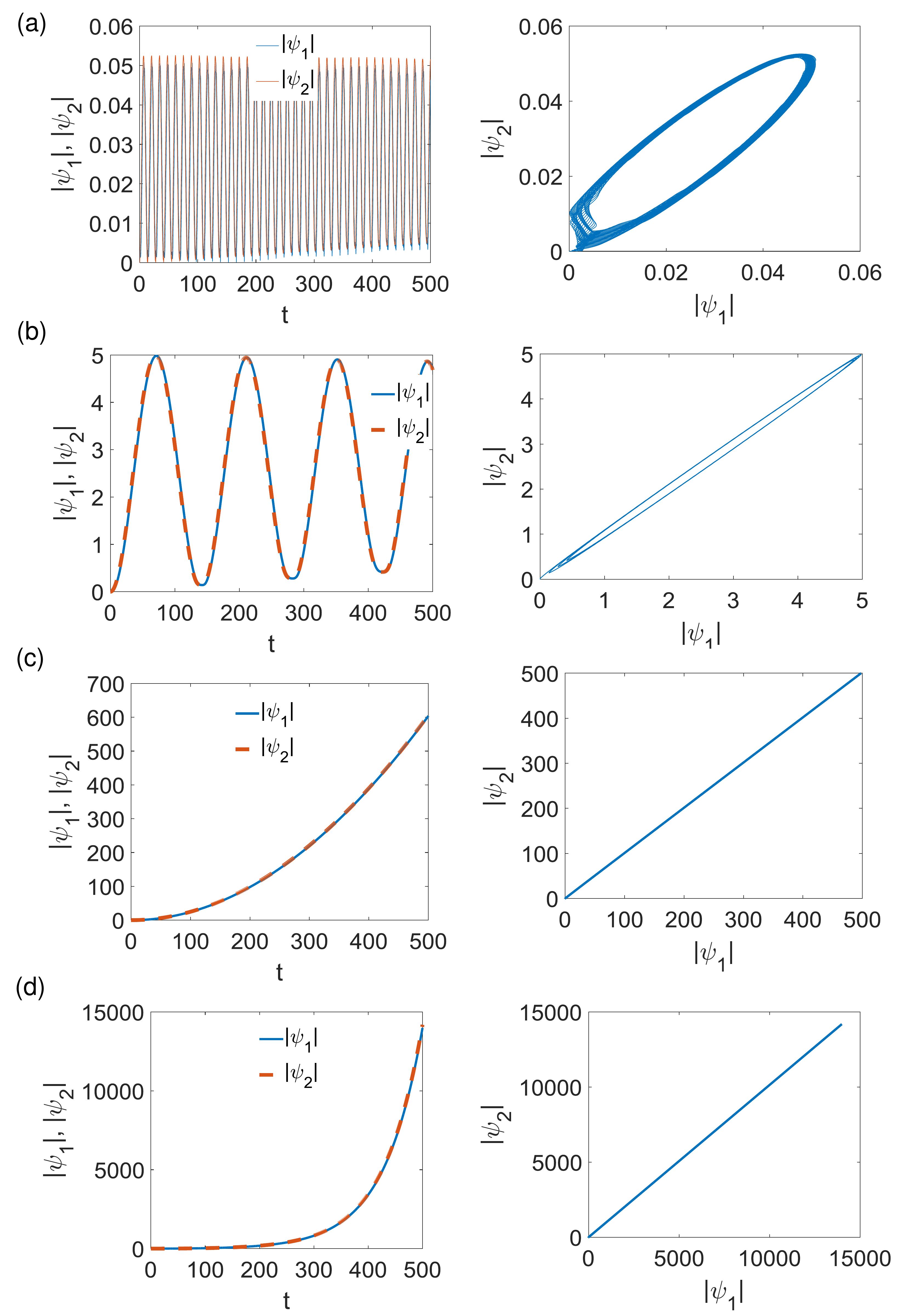}}
\caption{Evolution of field amplitudes: (a) $\kappa=1.1$. (b) $\kappa=1+10^{-3}$. (c) $\kappa=1$. (d) $\kappa=1-10^{-4}$.}
\end{figure}

In Fig. 8, we investigate the interplay between detuning and nonlinearity. The values of parameters are chosen wisely to get a better convergence from the numerical method. For instance, for larger detuning value, higher nonlinearity should be employed in order to compensate its effect, which can lead to divergence of the problem. 

The evolution of field amplitudes by time is shown in Fig. 8(a), for nonzero value of detuning $\epsilon=0.001$, while the other parameters are the same as in Fig. 7(b). Figure 8(b) shows that with adding an amount of nonlinearity $\chi_2=10^{-6}$, some resonances appear in the evolution path of the field amplitudes, and Fig. 8(c) shows that with $\chi_2=4.4\times10^{-5}$ a similar dynamic as in Fig. 7(b) (linear system with zero detuning) can be obtained, representing the compensation of detuning with nonlinearity. In Figure 8(d), we considered the coupling $\kappa=1$, which shows that the evolution of the system with ($\epsilon$,$\chi_2$)=($0.001,9\times 10^{-7}$) (green dash-dotted line) is almost identical with ($\epsilon$,$\chi_2$)=(0,0).

\begin{figure}
\centering
{\includegraphics[width=.49\textwidth]{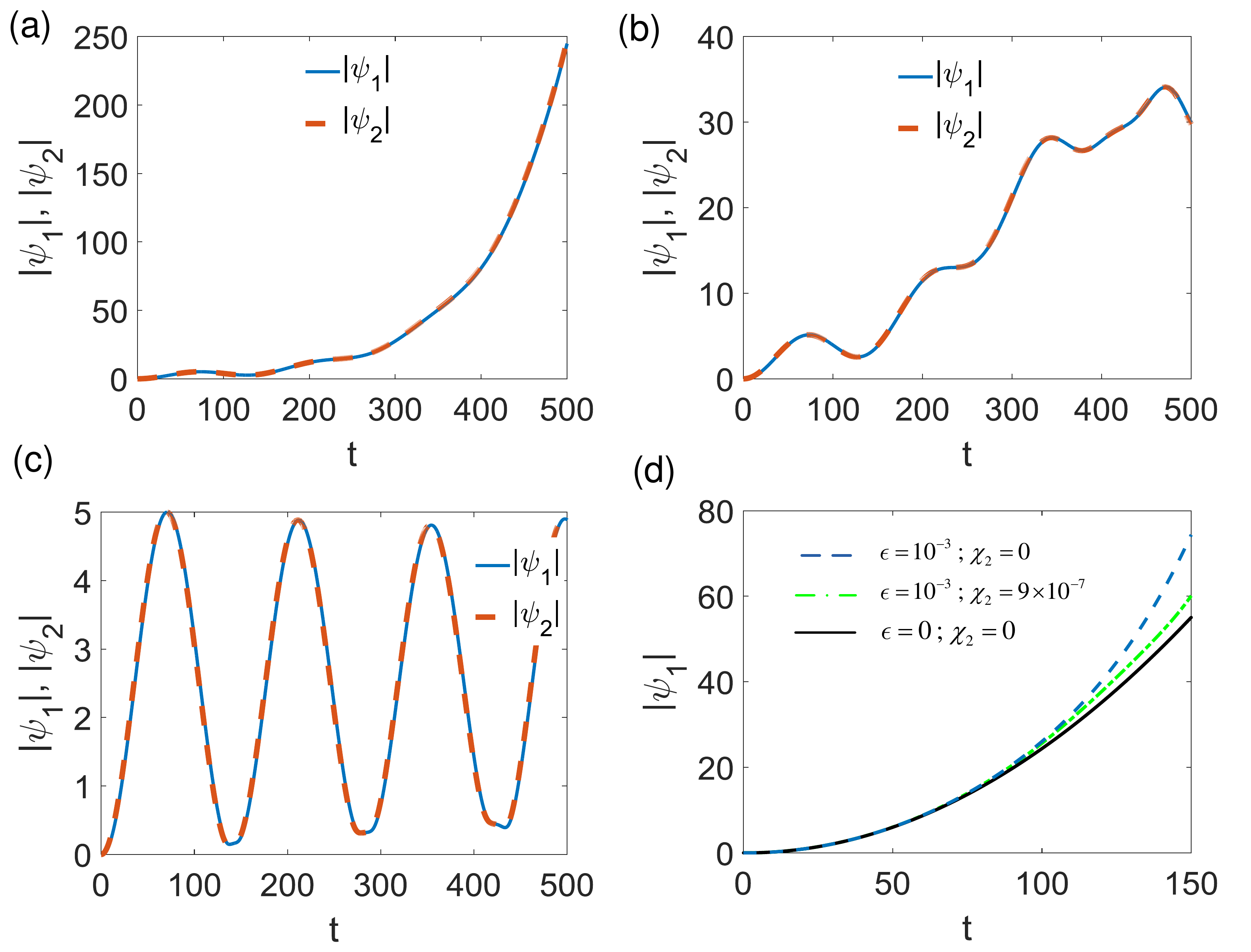}}
\caption{Evolution of field amplitudes for (a)-(c) $\kappa=1+10^{-3}$, $\epsilon=0.001$ and different values of nonlinearity, $\chi_2=0,\,10^{-6},\,4.4\times 10^{-5}$, and (d) $\kappa=1$ and different values of ($\epsilon$,$\kappa_2$).}
\end{figure}

\section{Conclusion}
 In conclusion, we studied a coupled resonator with gain and loss with a detuning parameter and nonlinearity for one of the resonators. The field mode is calculated by solving nonlinear coupled mode theory with self-consistent field and iteration methods. We have shown an excellent compensation of the detuning of the two-coupled cavity system by using an appropriate amount of nonlinearity. We have provided a full-wave numerical simulation to confirm the proposed approach. The method enables the system to work very close to the EP by tuning the intensity of the incident wave in a fast-process self-tuned scheme. 
\section*{ACKNOWLEDGMENTS}

\nocite{*}
\bibliography{EP_NL_Spectra}

\end{document}